# Is Zadeh's Least-Entered Pivot Rule Exponential?

Norman Zadeh[1]

**Abstract**

In 2011, Friedmann [F 7] claimed to have proved that pathological linear programs existed for which the Simplex method using Zadeh's "least-entered" rule [Z 14] would take an exponential number of pivots. In 2019, Disser and Hopp [DH 5] argued that there were errors in Friedmann's 2011 construction. In 2020, Disser, Friedmann, and Hopp [DFH 3,4] again contended that the least-entered rule was exponential. We show that their arguments contain multiple flaws. In other words, the worst-case behavior of the least-entered rule has not been established. Neither [F 7] nor [DFH 3,4] provides pathological linear programs that can be tested. Instead, the authors contend that their pathological linear programs are of the form "(P)" as shown on page 12 of [DFH 3]. The authors contend that "the constraints of (P) ensure that the probability of entering a vertex u is equal to the probability of exiting u." In fact, we note that the authors' constraints "(P)" are flawed in at least three ways: a) they require the probability of exiting u to exceed the probability of entering u, b) they require the probability of exiting some nodes to exceed 1, and c) they overlook flows from decision nodes to decision nodes.[2] At my request, in August of 2025, Disser, Friedmann, and Hopp provided me with their first ten purportedly pathological LPs and the graph of their first purportedly pathological Markov Decision Process (MDP1). It is shown that: a) their first two pathological LPs are infeasible if the variables are supposed to be probabilities, as the authors contend, and b) their first purportedly pathological LP does not match up with their first purportedly pathological MDP. In other words, the authors have not come close to providing counterexamples to the "least-entered" rule.

## 1   Introduction and background[3]

In 1972, Klee and Minty [KM 11] demonstrated the existence of linear programs with n inequality constraints in n non-negative variables that require $2^n - 1$ pivots when any improving column may enter and when the standard "max rate of increase" rule is followed.

Other constructions were soon after provided by Jeroslow [J 10] for a maximum increase rule, Zadeh [Z 13] for minimum cost flow algorithms, including the Simplex method with a "max rate of increase" rule, Avis-Chvatal [AC 1] for Bland's rule, Murty [M 12] and Fathi [F 6] for complementary pivot algorithms, and Goldfarb-Sit [GS 9] for a gradient selection rule. In 1980, Zadeh [Z 14] suggested the least-entered rule to prevent the Simplex method from taking an exponential number of pivots. That paper contained an offer of $1,000 to anyone who could find a sequence of pathological linear programs for which the least-entered rule was exponential.

---

[1] normanz@earthlink.net
[2] For the purposes of *understanding* the author's purported pathological LPs, it is convenient to think of their probability variable x(u,v) as a flow from node u to node v that is between 0 and 1.
[3] The author would like to thank Nimrod Meggido, David Avis, Ilan Adler, Richard Stone, and Anna Deza for their valuable input.

**The logic behind the "least-entered" rule**

In [Z 13], the trick used to generate the n+1st pathological transportation network from the nth pathological network was to add 2 nodes and 2n new arcs, and make the cost of the new arcs so high that all of the augmentations (pivots) in the nth network using a "max rate of increase rule" would be performed before any augmentations using the newly added arcs. Once all possible flow was sent from source to sink via the nth network, the next $2^n$ augmentations using the newly added expensive arcs reversed the $2^n$ augmentations that had occurred using arcs in the interior of the nth network. Each addition of 2 nodes thus doubled the number of augmentations (pivots).

From the perspective of Klee and Minty, the idea was to go through an exponential number of vertices in the nth polygon, add a few new dimensions, and then travel over all those same vertices (with the added dimensions) in the opposite direction.

<u>The least-entered rule complicates the inductive argument</u>

The least-entered rule makes any inductive argument substantially more difficult because the pivot sequence for the n+1st linear program cannot begin with the pivot sequence for the nth linear program. It must involve, early on, some of the newly added variables. For that reason, I believed that it could well be impossible to generate pathological examples for the "least-entered" rule.

## 2    Claimed pathological examples

In 2011, I received a phone call that Oliver Friedmann had solved "the Zadeh conjecture." Specifically, I was told that he had created a series of pathological linear programs for the "least-entered" rule. I also learned that he was giving a talk at UCLA. Since I had previously offered a $1,000 reward, I was asked to come by with a check. No paper or counterexamples were provided to me. Nevertheless, I paid the $1,000.

Prior to August 2025, I had not been aware of any counterexamples that I could examine and test.

## 3    Errors in the original 2011 papers

In July 2025, I discovered a 2019 paper written by Disser and Hopp [DH 5], which identified at least three flaws in Friedmann's 2011 claims, specifically in his 2011 paper [F 7]. On page 1 of [DH 5], it states, "We investigate Friedmann's lower bound construction and exhibit three flaws in his analysis: We show that (a) the initial policy for the policy iteration does not produce the required occurrence records and improving switches, (b) the specification of the occurrence records is not entirely accurate, and (c) the sequence of improving switches described by Friedmann **does not consistently follow Zadeh's pivot rule** (emphasis added).

In 2020, Disser, Friedmann, and Hopp joined forces to publish two additional papers [DFH 3, 4], which again claim that Zadeh's "least-entered" rule is exponential in the worst case.

## 4 Errors in the pathological LPs provided by Disser, Friedman, and Hopp

On August 8, 2025, I began communicating with the authors. On August 10, I received their first ten purported pathological linear programming formulations, the first of which is shown below and is referred to as LP1. They did not provide a starting basic feasible solution for any of their constructions. The $x_i$ variables below are supposed to be probabilities. (See page 12 of [DFH 3].)

**First Pathological Linear Program (offered by Friedmann, Disser, and Hopp)[4]**

$\alpha$ = 262145/524288, $\beta$ = 262143/524288, $\gamma$ = 1/262144

Maximize   $- 1024\, x_3 - 1024\, x_7 + 65536\, x_9 - 1024\, x_{10} + 256\, x_{11} + 69888\, x_{12}$

Subject to

$\alpha\, x_1 + x_2 - \beta\, x_3 - \beta\, x_5 - \beta\, x_7 - \beta\, x_{10} = 1$

$- x_2 + x_3 + x_4 = 1$

$- \beta\, x_1 - \beta\, x_3 + \alpha\, x_5 + x_6 - \beta\, x_7 - \beta\, x_{10} = 1$

$- x_6 + x_7 + x_8 = 1$

$x_9 + x_{10} - x_{11} = 1$

$- \gamma\, x_1 - \gamma\, x_3 - \gamma\, x_5 - \gamma\, x_7 - \gamma\, x_{10} + x_{11} + x_{12} = 1$

$x_i \geq 0$

**Problems with LP1**

**<u>Flaw 1:</u>**

There is no feasible solution for LP1 in which the variables are ≤ 1. In other words, there is no feasible solution for which the $x_i$ are probabilities. When the six constraints above are added together, noting from the above definitions that $\alpha - \beta = \gamma$ and $\alpha + \beta = 1$, we obtain

$x_4 + x_8 + x_9 + x_{12} = 6$

That implies that some $x_i$ must exceed 1 and therefore cannot be a probability.

---

[4] Disser and Friedmann provided me with two different objective functions, one of which switched the variables slightly ($x_4$ instead of $x_3$, $x_8$ instead of $x_7$, etc.). I have selected the objective function proposed by Friedmann. The objective function is irrelevant to this analysis.

**Flaw 2:**

LP2 is also infeasible if the $x_i$ are supposed to be probabilities. (See Appendix 1).

The authors assured me that their pathological LP constructions had the form shown on page 12 of [DFH 3], which I have reproduced below (hereinafter referred to as "(P)"). Friedmann used that form in his 2011 paper [F 7]. p(w,u) is the probability of moving from a randomization vertex w to a decision vertex u. Thus $0 \leq p(w,u) \leq 1$. (See page 10 of [DFH 3] or page 4 of [F 7]).

$$\max \sum_{(u,v) \in E_0} r(u,v) \cdot x(u,v)$$

$$\text{s.t.} \sum_{(u,v) \in E} x(u,v) \;-\; \sum_{\substack{(v,w) \in E_0 \\ (w,u) \in E_R}} p(w,u) \cdot x(v,w) = 1 \quad \forall u \in V_0 \qquad \text{(P)}$$

$$x(u,v) \geq 0 \qquad \qquad \forall (u,v) \in E_0$$

In other words, the authors contend that LP1 should have this form. The authors specifically state that the variables x(u,v) are probabilities. For example, on page 12 of [DFH 3], they state, "The variable x(u,v) for (u,v) $\in E_0$ represents the probability (or frequency) of using the edge (u,v). The constraints of (P) ensure that the probability of entering a vertex u is equal to the probability of exiting u." [F 7] and [DFH 4] have similar language.

**Flaw 3:**

The authors' first pathological MDP is inconsistent with their first pathological LP as well as their constraints "(P)."

Shown below is the authors' first purported pathological MDP (hereafter referred to as "MDP1"). They did not respond to my repeated requests to provide the full description of MDP1. However, the diagram does provide enough information to reveal multiple disparities between their MDP construction and their LP construction.

## First Pathological MDP Construction (offered by Friedmann, Disser, and Hopp)

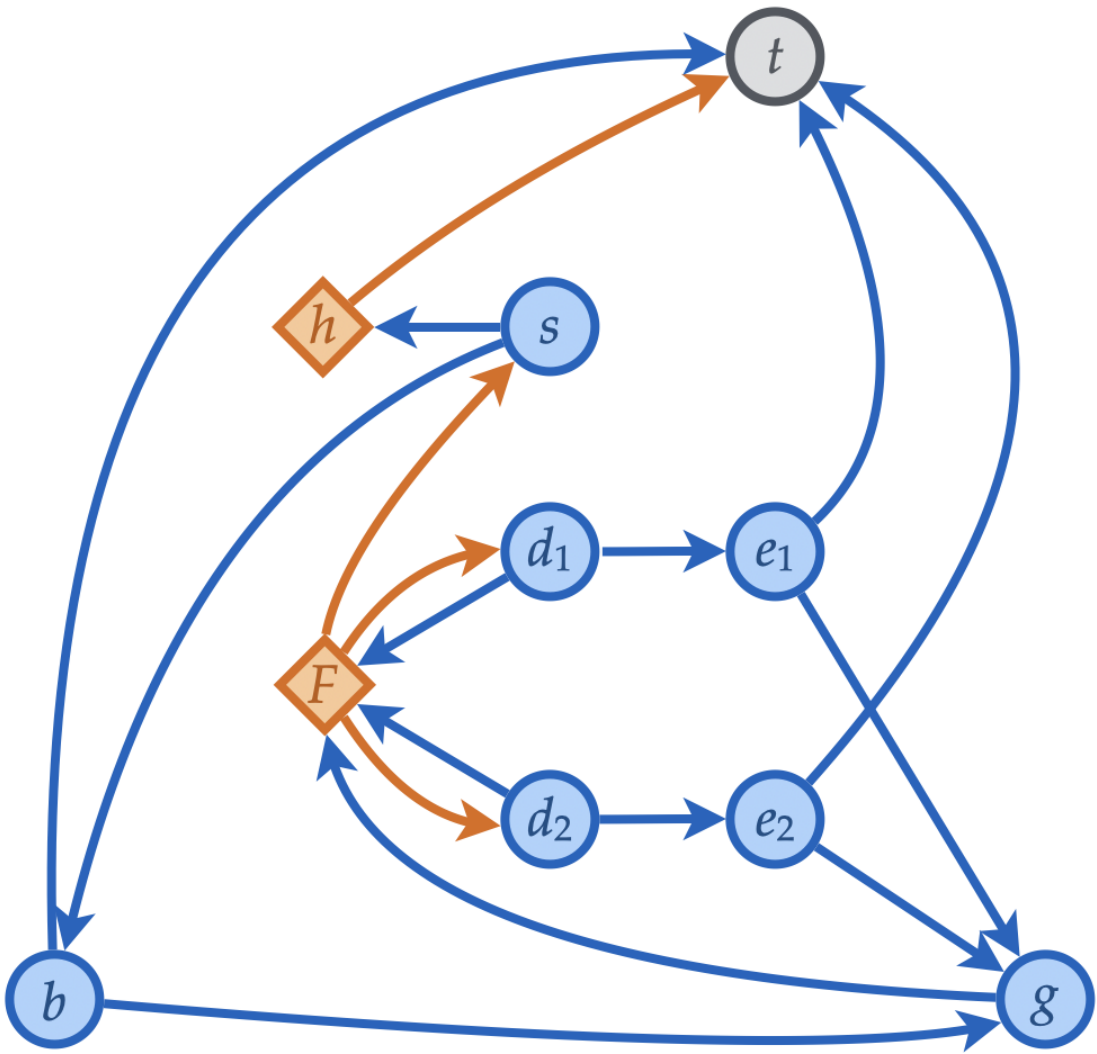

According to the authors, by writing down the flow conservation equations for MDP1 above, one is supposed to obtain the LP constraints "(P)" shown on page 12 of [DFH 3]. For example, on page 12 of [DFH 3], the authors write, "It is not difficult to see that the basic feasible solutions of (P) correspond directly to strategies of the MDP." They reference [AF 2][5].

However, when we write down the flow conservation equations for MDP1, we see that the formulations do not match up. For example, the conservation of flow constraint for node g in MDP1 is

$x(e_1,g) + x(e_2,g) + x(b,g) = x(g,F)$

Here, the left side is the flow (probability) into node g, and the right side is the flow out of node g. Equivalently, the constraint is

$x(e_1,g) + x(e_2,g) + x(b,g) - x(g,F) = 0$

Notice that the above equation is inconsistent with the constraints in LP1 or the constraints in "(P)" for at least three reasons. First, the right side is 0 as it should be in a conservation of flow equation. The right side in "(P)" as well as in LP1 is 1. Second, flows into node g are included from other decision nodes. Such terms are missing from "(P)," which only includes flows from randomization nodes into node u.

---

[5][AF 2] has the language, "It is not difficult to check that the BFS's of (P) correspond directly to the policies of the MDP."

Finally, the equations in LP1 have no constraints with four variables.

**Flaws in the authors' general LP construction "(P)"**

In the previous section, we touched on several flaws in the authors' general LP construction "(P)." We now discuss those flaws in more detail. First, the $\sum x(u,v)$ term on the far-left side of "(P)" is the probability of exiting u. The other summation term on the left side of "(P)" is the probability of entering u from any randomization vertex w. The authors have omitted a term corresponding to the probability of entering u from other decision nodes. Second, as written, "(P)" says that the probability of exiting vertex u is greater than 1 if the probability of entering u from a randomization vertex is positive. Probabilities cannot exceed 1. Finally, as written, "(P)" says that the probability of exiting vertex u exceeds the probability of entering vertex u by 1, which is contrary to what the authors claim "(P)" ensures.

It is noteworthy that Friedmann has not always had a "1" on the right side of "(P)." In at least one of his papers [see page 4 of FHZ 8], Friedmann has 0's on the right side.

I have repeatedly asked the authors the following questions and received no answers:

a) Should the coefficients on the right side of the constraints in "(P)" be 0?
b) Where did the 1's come from on the right side of their constraints?
c) Which node in MDP1 corresponds to the equation $- x_2 + x_3 + x_4 = 1$ in LP1?
d) Should the conservation of flow equation corresponding to node g in MDP1 be $x(e_1,g) + x(e_2,g) + x(b,g) = x(g,F)$?
e) Which constraint in LP1 corresponds to the conservation of flow equations for node g in the MDP1?

I received a response from Professor Disser on August 30, which said, "as far as I understand, it [MDP1] should have 13 actions and 7 player vertices. Accordingly, I would expect 13 variables and 7 constraints. I am not sure myself how this fits with the LPs that Oliver has sent. He'll have to answer this." LP1 has 12 variables and 6 constraints. I did not receive an answer from Oliver.

## 6   Summary

In [F 7], [DFH 3], and [DFH 4], the authors spend little time on their purported pathological linear programs. They provide no explicit examples or feasible starting solutions. Instead, they infer that their construction will take the form of "(P)" on page 12 of [DFH 3], a point they have confirmed to me directly. They claim the variables in these pathological LPs represent probabilities, and the constraints of "(P)" are meant to ensure that the probability of entering a vertex u equals the probability of exiting u. However, as written, the constraints "(P)" require that the probability of exiting u exceed the probability of entering u, which is contrary to what the authors claim. Also, the authors' constraints "(P)" ensure that the probability of exiting a vertex must exceed 1 in most instances. Probabilities cannot exceed 1. The authors' constraints "(P)" also overlook flows from decision nodes to decision nodes.

The authors' first two pathological LPs are infeasible if the variables are required to be probabilities. Their initial MDP and LP do not match up. The authors come nowhere close to demonstrating that pathological LPs exist that require an exponential number of pivots to solve using the least-entered rule.

**Appendix 1: Flaws in Friedmann's second purported linear programming pathology**

It is shown below that Friedmann's second purported linear programming pathology is infeasible if the variables are supposed to be probabilities, i.e., $x_i \leq 1$. Friedmann's second purported linear programming pathology (referred to as LP2) has 36 variables and 18 constraints. To establish infeasibility, it is only necessary to examine the first eight constraints, which are shown below.[6]

α = 15690529805/31381059609, β = 15690529804/31381059609, γ = 1/31381059609

**First eight constraints of LP2**

$α x_1 + x_2 - β x_5 - β x_{13} = 1$

$-x_2 + x_3 + x_4 = 1$

$-x_3 + x_5 + x_6 - x_{11} - x_{15} - x_{19} - x_{21} - x_{29} - x_{33} = 1$

$-x_4 + x_7 + x_8 - x_{12} - x_{16} - x_{20} - x_{22} - x_{30} - x_{34} = 1$

$- β x_6 + α x_9 + x_{10} - β x_{17} = 1$

$- x_{10} + x_{11} + x_{12} = 1$

$- β x_1 - β x_5 + α x_{13} + x_{14} = 1$

$- x_{14} + x_{15} + x_{16} = 1$

When the first eight constraints are added together, noting from the above definitions of α, β, and γ that α - β = γ and α + β = 1, we obtain

$γ x_1 + γ x_5 + α x_6 + x_7 + x_8 + α x_9 + γ x_{13} - β x_{17} - x_{19} - x_{20} - x_{21} - x_{22} - x_{29} - x_{30} - x_{33} - x_{34} = 8$

As defined above, γ is a tiny number, and α and β are close to 1/2. Even if every variable on the left side with a positive coefficient were 1, the left side would still be less than 8. In other words, the first eight constraints by themselves are infeasible.

---